\documentclass[sigplan,10pt,nonacm]{acmart}
\usepackage{subcaption}
\usepackage{multirow}
\usepackage{tikz}
\usepackage{mathalpha}
\usepackage{amsfonts}
\usepackage{tabularx}
\usepackage[export]{adjustbox}
\usepackage{url}
\usepackage{tablefootnote}
\usepackage{footnotehyper}
\usepackage{listings}
\usepackage{enumitem}
\makesavenoteenv{tabular}

\usepackage{mathtools}
\usepackage{pifont}
\usepackage{array,booktabs,ragged2e}

\AtBeginDocument{%
  \providecommand\BibTeX{{%
    \normalfont B\kern-0.5em{\scshape i\kern-0.25em b}\kern-0.8em\TeX}}}

\setcopyright{acmcopyright}

\begin{document}

\newcommand{\fixme}[1] {{\bf [FIXME: #1]}}
\newcommand{\zeroploit}[0]{{\it Zeroploit}}
\newcommand{\autoploit}[0]{{\it AZP}}
\newcommand{\avgBucketSpeedup}[0]{16.4\%}
\newcommand{\avgBestOfBucketSpeedup}[0]{18\%}
\newcommand{\fpsSpeedup}[0]{3.5\%}
\newcommand{\fpsBestOfSpeedup}[0]{3.9\%}
\def\checkmark{\tikz\fill[scale=0.4](0,.35) -- (.25,0) -- (1,.7) -- (.25,.15) -- cycle;} 

\newcommand{\app}[1]{{\tt#1}}

\newcommand*{\matr}[1]{\mathbf{#1}}
\newcommand{\eleminvalid}[0]{$X$}
\newcolumntype{R}[1]{>{\RaggedLeft\arraybackslash}p{#1}}

\newcommand{\nvidia}[0]{NVIDIA}
\newcommand{\rtx}[0]{RTX\texttrademark}
\newcommand{\pascal}[0]{Pascal\texttrademark}
\newcommand{\gpuname}[0]{GeForce \rtx\ \gpu}
\newcommand{\gpu}[0]{2080}

\title{\autoploit: Automatic Specialization for Zero Values in Gaming Applications}


\author{Mark W. Stephenson}
\email{mstephenson@nvidia.com}
\affiliation{%
  \institution{NVIDIA}
  \city{Austin}
  \state{TX}
  \country{USA}
}

\author{Ram Rangan}
\email{rrangan@nvidia.com}
\affiliation{%
  \institution{NVIDIA}
  \city{Bangalore}
  \state{Karnataka}
   \country{India}
}



\begin{abstract}

Recent research has shown that dynamic zeros in shader programs of gaming applications can be effectively leveraged with a profile-guided, code-versioning transform. This transform duplicates code, specializes one path assuming certain key program operands, called versioning variables, are zero, and leaves the other path unspecialized. Dynamically, depending on the versioning variable's value, either the specialized fast path or the default slow path will execute. Prior work applied this transform manually and showed promising gains on gaming applications.  In this paper, we present \autoploit{}, an automatic compiler approach to perform the above code-versioning transform. Our framework automatically determines which versioning variables or combinations of them are profitable, and determines the code region to duplicate and specialize (called the versioning scope). \autoploit\ takes operand zero value probabilities as input and it then uses classical techniques such as constant folding and dead-code elimination to determine the most profitable versioning variables and their versioning scopes. This information is then used to effect the final transform in a straightforward manner.  We demonstrate that \autoploit\ is able to achieve an average speedup of \avgBucketSpeedup\ for targeted shader programs, amounting to an average frame-rate speedup of \fpsSpeedup\ across a collection of modern gaming applications on an \nvidia\ \gpuname\ GPU.

\end{abstract}

\maketitle

\section{Introduction}
\label{sec:intro}

Graphics processing units~(GPUs) have enjoyed consistent generation-over-generation performance growth since their inception, thanks in large part to 2D CMOS scaling. However, as GPU performance growth fueled by Moore's Law comes to an end, the gaming industry is in urgent need of novel architectural and software solutions to deliver similar performance growth to sustain innovations in our quest for photo-realistic real-time rendering. 

To that end, Rangan et al. recently introduced \zeroploit, a profile-guided optimization to exploit dynamically zero valued operands in shader programs of gaming applications~\cite{rangan:20:taco}. \zeroploit\ uses the knowledge that one of the input operands of a multiply (or a similar) operation is mostly zero dynamically to rearrange the code so as to avoid computing the {\em other} source operand whenever the former is zero, since anything multiplied by a zero results in a zero.  Though this is not safe as per IEEE 754 rules~\cite{ieeefp:2019} since the {\em other} source operand could evaluate to a floating point {\em NaN},  $-\infty$ or $+\infty$, game developers typically allow for such IEEE-unsafe floating point optimizations\footnote{In Microsoft's high-level shading language~(HLSL), this is achieved by explicitly setting the {\em refactoringAllowed}\cite{msdnRefactorFlags} global flag and dropping the {\em precise} storage class specifier for variable declarations~\cite{hlslvariabledcl}.} to  opportunistically squeeze out additional performance as well as to ensure NaN values do not leak into render targets. \zeroploit\ leverages this developer-granted permission to transform a code region into a zero-specialized fast path and an unspecialized slow path. In the aforementioned work, \zeroploit\ was applied manually to graphics shader programs, both in terms of opportunity detection as well as the code-versioning transform itself. In this paper, we strive for an automatic compiler solution for the \zeroploit\ optimization for shader programs of gaming applications.

\begin{figure}
\begin{subfigure}{.49\linewidth}
\centering
\begin{footnotesize}
\begin{verbatim}
// r1 == 0, 50% of the time
1. r1 = expressionAvblEarly()  
2. r0 = expensiveWork1() 
3. r2 = r0 x r1                
4. r3 = r2 x expensiveWork2()
5. r4 = r3 + r5
\end{verbatim}
\end{footnotesize}
\caption{Original.}
\end{subfigure}%
\begin{subfigure}{.49\linewidth}
\centering
\begin{footnotesize}
\begin{verbatim}
1. r1 = expressionAvblEarly()
2. if (r1 == 0) { // fast path
3.  // specialization hint 
4.  r1 = 0  
5.  r0 = expensiveWork1()
6.  r2 = r0 x r1
7.  r3 = r2 x expensiveWork2()
8. } else {  // slow path
9.  r0 = expensiveWork1()
10. r2 = r0 x r1
11. r3 = r2 x expensiveWork2()
12.}
13.r4 = r3 + r5
\end{verbatim}
\end{footnotesize}
\caption{After \zeroploit.}
\end{subfigure} 
\begin{subfigure}{\linewidth}
\centering
\begin{footnotesize}
\begin{verbatim}
1. r1 = expressionAvblEarly()
2. if (r1 == 0) { // fast path
3.  r3 = 0
4. } else {  // slow path
5.  r0 = expensiveWork1()
6.  r2 = r0 x r1
7.  r3 = r2 x expensiveWork2()
8. }
9. r4 = r3 + r5
\end{verbatim}
\end{footnotesize}
\caption{After \zeroploit\ + specialization.}
\end{subfigure}
\caption{Example illustrating \zeroploit. Here, {\em r1} is the versioning variable. {\em expensiveWork1()} and {\em expensiveWork2()} get specialized away due to \zeroploit's backward and forward slice specialization respectively. Instructions on lines 2, 3, and 4 of (a) form the versioning scope and get duplicated. One of the versions gets specialized to form the fast path, while the other remains unspecialized and serves as the default, slow path.}
\label{fig:simpleexample}
\vspace{-0.1in}
\end{figure} 

Figure~\ref{fig:simpleexample} illustrates \zeroploit. It shows the two basic requirements of the \zeroploit\ transform, namely, the identification of a versioning variable and the identification of a versioning scope. In its basic form, a versioning variable is an operand that a value profiler identifies as having a high probability of being zero dynamically (e.g., the output of a saturating arithmetic operation). The versioning scope is the region of the input code that gets duplicated to create the specialized fast path and unspecialized slow path code versions. 

An automatic compiler solution to \zeroploit\ must solve three key challenges. First, given that shader programs typically have fat and complex data flow graphs~(DFGs) and the probability of their operands being dynamically zero is more than 11\%~\cite{rangan:20:taco}, automatic identification of versioning variables and versioning scopes requires the ability to carefully consider, rank, and choose from among multiple potential versioning variables, each with their own zero probabilities and versioning scopes that may overlap. Besides, large versioning scopes can lead to instruction cache thrashing and the ranking algorithm must appropriately trade this off with specialization benefits. Second, GPUs' mixed scalar-vector instruction set architectures~(ISAs) require analysis and transform intelligence to optimally specialize concurrent DFGs originating at vector instructions (e.g. texture load). Likewise, a versioning condition based on all components of a vector load being zero will need to automate the creation of a composite or combined versioning variable prior to effecting the transform. Finally, an automatic solution must ensure that control flow divergence due to code versioning does not adversely impact performance on single-instruction multiple-threads~(SIMT) GPUs.

{\bf Contributions:} In this paper, we present \autoploit, an automatic technique for the \zeroploit\ optimization that effectively addresses the above challenges. Leveraging analysis techniques from classical optimizations such as constant folding and dead-code elimination, we devise novel heuristics to estimate benefit, rank, and identify profitable versioning variables and versioning scopes in practically linear time. We present techniques to effectively address challenges arising from mixed vector-scalar ISAs. \autoploit's transform relies on SIMT-wide {\em vote}~\cite{ptx:19} or equivalent operation to ensure that versioning branches are dynamically convergent. We describe the \autoploit\ compiler pipeline in detail and show how the above solutions fit together within this framework. 

We evaluate \autoploit\ on a suite of modern, heavily-optimized gaming applications and show that the automatic approach presented in this paper can provide an average speedup of \avgBucketSpeedup\ for targeted shaders on an \nvidia\ \gpuname\ GPU, which translates to an average frame-rate speedup of \fpsSpeedup. We show that \autoploit's compile-time overhead is suitable for just-in-time environments such as GPU drivers generating code for interactive gaming applications.

\section{Background}
\label{sec:background}

This section provides a high-level introduction to three topics: graphics programs, execution model in \nvidia\ GPUs, and profile-guided optimizations. 

\subsection{Graphics Programs}
Modern gaming applications are most popularly developed in Direct3D 11~\cite{d3d11}, Direct3D 12~\cite{d3d12}, OpenGL~\cite{opengl}, and Vulkan~\cite{vulkan} APIs. Without exception, all graphics APIs use a two-level hierarchy to convey work from the CPU to the GPU: an API layer and a shader program layer. The API layer is used to setup GPU state (e.g., enable/disable depth testing, enable/disable blending, etc.); bind resources such as constant buffers, textures (read-only input data), render targets (write-only output data), or general multi-dimensional arrays called unordered access views (UAVs, which are read-write); and bind shader programs for use in GPU work calls (which could be graphics \emph{draw} calls or compute \emph{dispatch} calls). A GPU work call typically causes one or more threads of a corresponding bound shader program to be run. Shader programs are typically written in high-level languages such as the high-level shading language~(HLSL) used with Direct3D APIs~\cite{hlsl}, or the OpenGL Shading Language~(GLSL) used with OpenGL or Vulkan APIs~\cite{glsl}. 

A typical API sequence interleaves various state setup commands with GPU work calls. Several such API commands can be in flight at any given point in the GPU. State setup commands take very little time and most of the GPU time is  spent in GPU work calls. A single frame of a modern game typically requires hundreds or even thousands of state setup and GPU work calls.  
The hundreds or thousands of API calls that make up a frame can exhibit vastly different performance characteristics. An API call can be limited by the performance of fixed function units, CPU-GPU data transfer latencies, or the performance of programmable shaders, whose performance can in turn be limited by memory bandwidth, instruction issue rate, or latency of memory loads.  \autoploit\ is applicable only to the portion of a gaming application's frame time that is dominated by programmable shaders, namely \emph{pixel} and \emph{compute} shaders, which are the two most expensive shader types.

\subsection{Execution Model in \nvidia\ GPUs}

Using \nvidia's terminology, shader programs execute on programmable processing cores called streaming multiprocessors~(SMs). 
Threads of shader programs are grouped into convenient physical entities called {\em warps}. A warp can span {\em at most} 32 threads. Instruction fetch, decode, and scheduling in the SM happens at a warp granularity, while operand fetch, execution in functional units, and register writeback happen at a per-thread granularity.  SM execution is most efficient when handling full warps, since all its functional units are fully utilized. The SM implements a single-instruction, multiple threads~(SIMT) execution model~\cite{ptx:19} whereby, each thread of a warp maintains its own program counter~(PC). If threads of a warp diverge, i.e. execute at different PCs, SM datapath utilization suffers. To mitigate divergence the SM uses joint hardware-software mechanisms to make diverged threads re-converge at well-defined ``synchronization" points in the program.  We will build on the above background when discussing \autoploit's heuristics later in the paper. 

\subsection{Profile-guided optimizations}

Profile-guided optimizations~(PGOs) have historically focused on collecting control flow profiles to determine the execution weights of various blocks of code. Compiler passes such as inlining, register allocation, predication, loop unrolling, etc. then make use of accurate information about execution weights to generate suitably optimized code. For example, a compiler can layout code based on profile feedback to improve instruction cache performance by packing frequently taken blocks closer to each other. Such execution weight based PGOs have proven effective in a variety of programs, from general purpose programs~\cite{cohn:99:wfdo} to Web browsers~\cite{chromepgo}. Many modern commercial compilers support PGOs~(e.g.~\cite{mscompiler, intelcompiler}).

PGOs typically involve compiling and executing code in two different ways. First, the baseline code needs to be instrumented for profile collection and profiles collected. Next, a second compilation may then use the collected profile information to transform the baseline code suitably to effect a PGO. At a high level, the \autoploit\ optimization presented in this paper uses a similar dual compilation approach.  \autoploit\ differs from typical PGOs in its use of value profiles, instead of control flow profiles, to optimize code.  A more detailed comparison with prior value specialization work is given in Section~\ref{sec:related}.

\newcommand{\verstart}[0]{{\em ver\_start}}
\newcommand{\verend}[0]{{\em ver\_end}}

\subsection{Zeroploit}
\label{sec:zeroploit}

We now provide background information and terminology related to \zeroploit~\cite{rangan:20:taco}. Rangan et al. define a {\em zero-propagator} as an instruction that produces zero as output due to one or more of its source operands being zero (e.g., a multiply operation) and a {\em zero-originator} as an instruction that produces zero as output from non-zero inputs, due to its semantics (e.g., a saturating arithmetic operation). They further define a {\em useful zero-originator} as a zero-originator whose output is consumed by a zero-propagator.  

A typical \zeroploit\ opportunity has three main ingredients, namely: 
\begin{enumerate}[wide, labelwidth=!, labelindent=0pt]
\item A program variable that is produced by a useful zero-originator and is consumed by a zero-propagator. This variable (or its equivalent) will become the {\em versioning variable} in the transformed code, based on whose dynamic value, execution will jump to either a fast path or a slow path. 
\item The {\em other} operand of the zero-propagator is expensive to compute.
\item The expression computing the versioning variable is available early. At a minimum, this expression must be independent of the {\em other} operand. In general, if the backward slices of the versioning variable and the {\em other} operand intersect farther up a shader's program dependency graph (or do not intersect at all), the better it is for versioning scope formation as it can allow for larger versioning scopes and thus more specialization. 
\end{enumerate}

We refer to the example in Figure~\ref{fig:simpleexample} here, as we describe the steps of the \zeroploit\ transform below. The code in Figure~\ref{fig:simpleexample}a, which is assumed to satisfy all three prerequisites, can be \zeroploit-transformed to the code in Figure~\ref{fig:simpleexample}b as follows.  First, \zeroploit\ identifies a set of operations that will be affected by either forward or backward slice (including recursive backward slice) specialization based on the versioning variable being zero. The region of code from the first affected operation to the last forms the versioning scope. Next, it duplicates operations in the versioning scope. One of the copies or versions, which will become the specialized, fast path, is prefixed with an explicit assignment of zero to the versioning variable to enable the compiler back-end to easily notice the specialization opportunity in that scope and apply classical optimizations such as constant folding, constant propagation, and dead code elimination to instructions in that path. The other version will serve as the default, unspecialized, slow path. Finally, a conditional branch predicated on the versioning variable being equal to zero is added to dynamically steer execution to the fast path or the slow path. Figure~\ref{fig:simpleexample}c shows the final optimized code after the fast path has been specialized by a compiler back-end.  
As identified in Section~\ref{sec:intro}, an automatic \zeroploit\ solution for gaming applications running on SIMT GPUs must solve three key challenges. It must be able to evaluate, rank, and choose from among several potential versioning variables, deal with challenges posed by mixed scalar-vector ISAs when determining versioning variables and scopes, and ensure SIMT warp divergence penalties do not negate specialization benefits. 
The next section describes how we address the above challenges with \autoploit.

\section{\autoploit}
\label{sec:autoploit}

We now describe \autoploit, which automatically transforms shader programs to profitably exploit likely zero operands. This section discusses how our compiler identifies candidate versioning variables, our zero-value profiler, and the compiler analyses and transformations \autoploit\ performs.  In the remainder of the paper when we generically say that the compiler specializes on a versioning variable, we implicitly mean that the compiler specializes a region of code in which the variable is guaranteed to be \emph{zero}, and it guards that region with a runtime check that invokes either the specialized region, or the default \emph{fallback} region. Without loss of generality we assume the compiler's intermediate representation (IR) is in static single assignment (SSA) form.

\begin{figure}
\includegraphics[width=\linewidth]{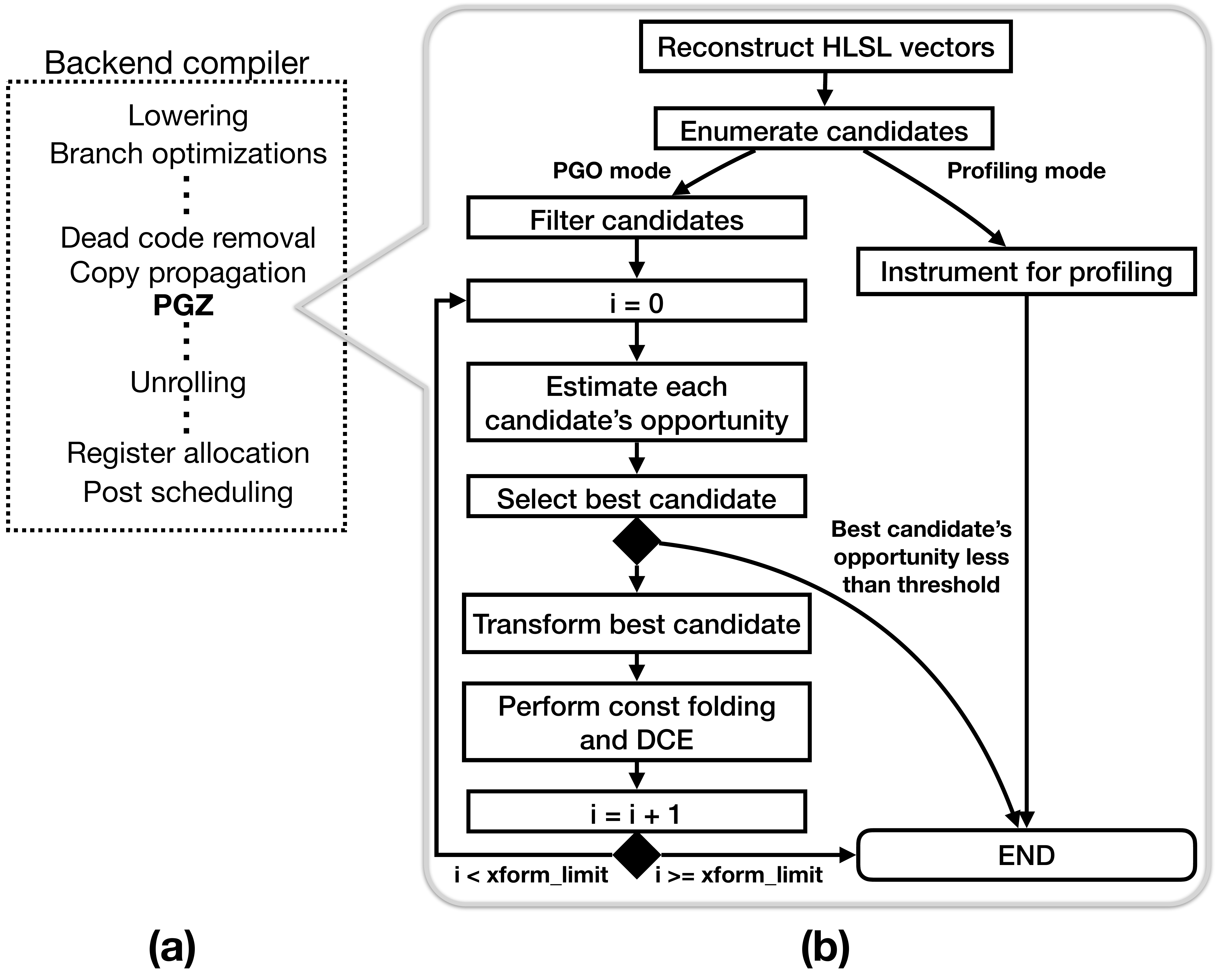}
\caption{\autoploit\ flow.  \autoploit\ runs early in our prototype's compiler flow as (a) indicates, before many of our compiler's significant passes run.  As (b) shows, after enumerating the potential candidates, \autoploit\ runs in one of two modes: instrumentation, or PGO.  The instrumentation path (labelled \emph{profiling mode}) is for collecting the offline profiles that drive the \autoploit\ optimization (labelled \emph{PGO mode}).  The PGO path iteratively evaluates and transforms candidates.  In each iteration of the evaluate-transform loop, \autoploit\ estimates the effectiveness of transforming each candidate, and greedily transforms the candidate with the best estimated opportunity.  The transformation may expose new opportunities on the next iteration. }
\label{fig:zp_compiler_flow}
\vspace{-0.075in}
\end{figure}

Figure~\ref{fig:zp_compiler_flow}(a) shows where \autoploit\ runs with respect to other compiler phases.  Our profiling pass, as well as the \autoploit\ analyses and transformations run early in our backend compiler's flow, and as a result the IR that \autoploit\ sees is far from optimal. However, by running \autoploit\ {\em after} classical optimizations like dead code elimination and copy propagation, we ensure \autoploit\ operates on code without any static redundancy, which reduces the number of potential operands it needs to profile and evaluate. We choose to run \autoploit\ earlier than advanced backend optimizations because loop optimizations such as unrolling and software pipelining can increase the number of potential versioning variables our optimization has to consider. As with most compiler optimizations we recognize that \autoploit\ is sensitive to phase orders. Future work will consider alternate phase orders.

\subsection{Identifying candidate versioning variables}

Regardless of \autoploit's ordering with respect to other compiler passes, \autoploit's first task is to enumerate the set of possible versioning variables in a shader, which we henceforth refer to as \emph{candidates}.  We call them candidates because at this point in \autoploit's flow the compiler does not know the potential upside to specialization on any of the versioning variables. The candidate enumeration step of Figure~\ref{fig:zp_compiler_flow}(b) proceeds as follows:

\noindent
{\bf 1. Find scalar candidates}: \autoploit\ makes the distinction between \emph{scalar} candidates and \emph{vector} candidates.  Scalar candidates are trivial to enumerate since they are merely the virtual register writes in a program.  To reduce the optimization search space \autoploit\ does not consider as candidates constant literal moves because traditional compiler transformations such as constant propagation and folding will likely optimize these away; nor does \autoploit\ consider register-to-register moves as candidates because they simply rename the move's source operand and are therefore redundant.  All other register writes are potential scalar candidates.  In SSA form there is a one-to-one correspondence between candidates and virtual registers.

\noindent
{\bf 2. Find vector candidates}:  In addition to scalar candidates, \autoploit\ groups subsets of scalar candidates together to form \emph{vector} candidates.  In fact, many of \autoploit's biggest gains come from jointly specializing multiple scalar candidates in the IR, which is also supported by prior observations~\cite{rangan:20:taco}.  We have seen cases where jointly specializing on the conjunction of two or more scalar candidates exposes specialization opportunities, whereas specializing on any scalar candidate in isolation exposes none. Blindly considering arbitrary combinations of candidates is combinatorial in the number of versioning variables and is therefore not tractable.  To gain insights into what combinations \autoploit\ should consider we manually inspected dozens of shader programs and found that many specializable combinations of variables are \emph{explicitly} combined by the programmer into short vectors, typically into RGBA or XYZW components. Manual inspection of shader program code led to the insight that some specialization opportunities require whole vectors to be zero.  So while \autoploit\ does not consider packing arbitrary combinations of candidates, we found that packing candidates that are elements of a programmer-specified vector into a \emph{vector} candidate is sufficient to expose additional specialization opportunities.  

A challenge we face when attempting to pack candidates into their programmer-specified vectors is that short vector information does not propagate from the programming language to the compiler's backend IR.  In fact, for DirectX 12, the explicit vectors in the high level shader language (HLSL) are lost before backend compilation begins.  To reconstruct programmer-specified vectors we leverage prior art on compiling for short, single instruction multiple data (SIMD) instruction sets.  In particular, for this work we implemented portions of Larsen and Amarasinghe's ``SLP" algorithm for packing independent and isomorphic instructions into SIMD \emph{superwords}~\cite{slp}.  We note that for our purposes since we are not actually packing candidates into wide registers nor scheduling instructions to execute in SIMD fashion, a best-effort implementation that forms natural groups of candidates suffices.  \autoploit\ creates a new set of vector candidates from the superword-level vectors formed by applying SLP~\cite{slp}.

\noindent
{\bf 3. Enumerate candidates}: \autoploit\ enumerates the union of scalar candidates ($S$) and vector candidates ($V$).  That is, \autoploit\ assigns each candidate with an identifier that it uses to associate candidates with runtime profiles.  

Formalizing our discussion of candidates, each scalar candidate $s \in S$ corresponds to exactly one SSA virtual register definition ${vr}_s$; and a virtual register ${vr}$ associates with at most one scalar candidate, $s$.  Virtual registers can also correspond to a vector candidate $v \in V$ \emph{and} a scalar candidate $s \in S$. Each vector candidate represents a set of virtual registers, e.g., ${vr}_v = \{ {vr}_x, {vr}_y, {vr}_z, {vr}_w \}$ where $x, y, z, w \in S$ and are distinct.  We limit the number of virtual registers associated with a vector candidate to four.

As Figure~\ref{fig:zp_compiler_flow}(b) shows, after candidate enumeration \autoploit\ can take one of two distinct paths.  On one path \autoploit\ instruments the shader to collect runtime profiles, and on the other path \autoploit\ ingests a runtime profile and attempts to specialize the shader for likely zeros.  We first discuss program instrumentation for profiling.

\subsection{Instrumenting for profiling}

For each candidate enumerated above, our profiler records the likelihood that the candidate is zero at runtime. Rangan et al.~present a general value profiling framework~\cite{rangan:20:taco} that builds upon the ideas presented in~\cite{calder:97:micro}.  However, they also note that they did not see significant opportunities to leverage values other than zero.  Applying this insight, instead of implementing a general value profiler, we designed a simple profiler that discovers the likelihood that a candidate is zero.  Our profiler instruments each scalar candidate $s \in S$ to record both the total number of writes, ${writes}_s$, to the candidate's associated virtual register ${vr}_s$ in addition to the number of those writes that were zero, ${zero}_s$.  Our zero-value profiler leverages \nvidia's global memory atomic increment operations for updating these counters. 

For each scalar candidate $s$, we can easily estimate the likelihood, $P({vr}_s = 0)$, that the associated versioning variable is zero according to $P({vr}_s = 0) \approx \frac{{zero}_s}{{writes}_s}$.  Vector candidates, on the other hand require a different approach.  Ideally we would estimate the likelihood that all elements of the vector are jointly zero, e.g., $P({vr}_v = 0) = P({vr}_x = {vr}_y = {vr}_z = {vr}_w = 0)$.  A profiler could determine the joint likelihood by only incrementing ${writes}_v$ after all elements of the vector are written, and by only incrementing ${zero}_v$ if all elements of the vector are zero.  Alternatively, and the approach we use in our prototype, is to estimate $P({vr}_v = 0)$ assuming independence among the individual candidates, i.e.,
\begin{align*}
P({vr}_x = 0) \cdot P({vr}_y = 0) \cdot P({vr}_z = 0) \cdot P({vr}_w = 0) 
\end{align*}

The final profiling decision we discuss relates specifically to SIMT execution.  If the versioning variable is not \emph{uniformly} zero across the threads of a warp, should we still consider specializing code on that variable?  \autoploit\ only considers warp-uniform zero specialization primarily because it is the \emph{conservative} choice.  One of the primary metrics GPU programmers use to gauge the performance of their code is \emph{SIMT efficiency}, i.e., how many threads in a warp are simultaneously active on average.  Were \autoploit\ to introduce \emph{divergent} branches, where the guarding runtime specialization check causes the warp to \emph{serially} execute the specialized region \emph{and} the fallback region), it would decrease SIMT efficiency and quite likely reduce performance.\footnote{The authors of \zeroploit\ note that it is still beneficial to introduce divergent branches when the shader is limited by texture throughput and specialization removes texture operations~\cite{rangan:20:taco}.  Such scenarios are difficult for tools to identify without hardware performance monitor feedback.} For this reason, \autoploit's profiler only increments a candidate $c$'s zero counter ${zero}_c$ when ${vr}_c$ is zero for all threads in the warp, by using the {\em vote.all} flavor of \nvidia's {\em vote} instruction to check for warp-wide convergence~\cite{ptx:19}.\footnote{The {\em vote.all} instruction returns {\em True} if a Boolean predicate is {\em True} across all active threads of a SIMT warp and {\em False} otherwise.} 

\subsection{Profile guided zero specialization}

We now discuss the steps our prototype performs to specialize a shader. As shown in Figure~\ref{fig:zp_compiler_flow}(b), when the driver supplies the backend compiler with a valid runtime profile, \autoploit\ attempts to specialize a shader to exploit likely zeros.  The basic flow of our profile-guided specialization is to iteratively estimate each candidate's performance potential and then transform the best candidate if it is above a given threshold. We now describe each step of this flow.

\subsubsection{Filtering candidates}

As we soon discuss, each step of the estimate-transform loop shown in Figure~\ref{fig:zp_compiler_flow}(b) is expensive.  We can significantly improve the compile-time overhead by filtering out candidates where $P({vr}_c = 0) < P_{Thresh}$.  In Section~\ref{sec:eval} we show how $P_{thresh}$ affects compile time, but for all other results we present, $P_{thresh} = 0.32$.  For vector candidates, we remove any element from the candidate if its zero likelihood is less than $P_{thresh}$.  We can also skip the evaluation of any zero-propagator candidates $zp$ with $P({vr_{zp}} = 0)$ if one of its zero-propagating inputs is generated by a candidate $c$ with $P({vr_c} = 0) >= P({vr_{zp}} = 0)$.
This follows because in this scenario, specializing for candidate $c$ will also specialize candidate $zp$. Eliminating zero-propagator candidates in this fashion through a transitivity analysis helps whittle down the list of candidates further to just root zero-originators.

\subsubsection{Estimating candidate opportunity}

\autoploit\ introduces control flow, which not only bloats code, but also has the potential to negatively impact scheduling by splitting independent long latency instructions across basic blocks and artificially serializing them.  Therefore, we need to balance the deleterious effects of \autoploit's code transformation with its potential upside. With a set of candidates to consider, \autoploit\ estimates the benefit of each candidate.  Our algorithm greedily selects the candidate with the best estimated opportunity, and transforms the shader to specialize for the candidate's associated variable(s).  While the actual specialization transformation is straightforward, estimating a candidate's benefits \emph{quickly} is more difficult. Note that shaders commonly have hundreds of candidates, so we must limit the compile-time complexity of opportunity estimation.

\begin{figure}[t]
\includegraphics[width=\linewidth]{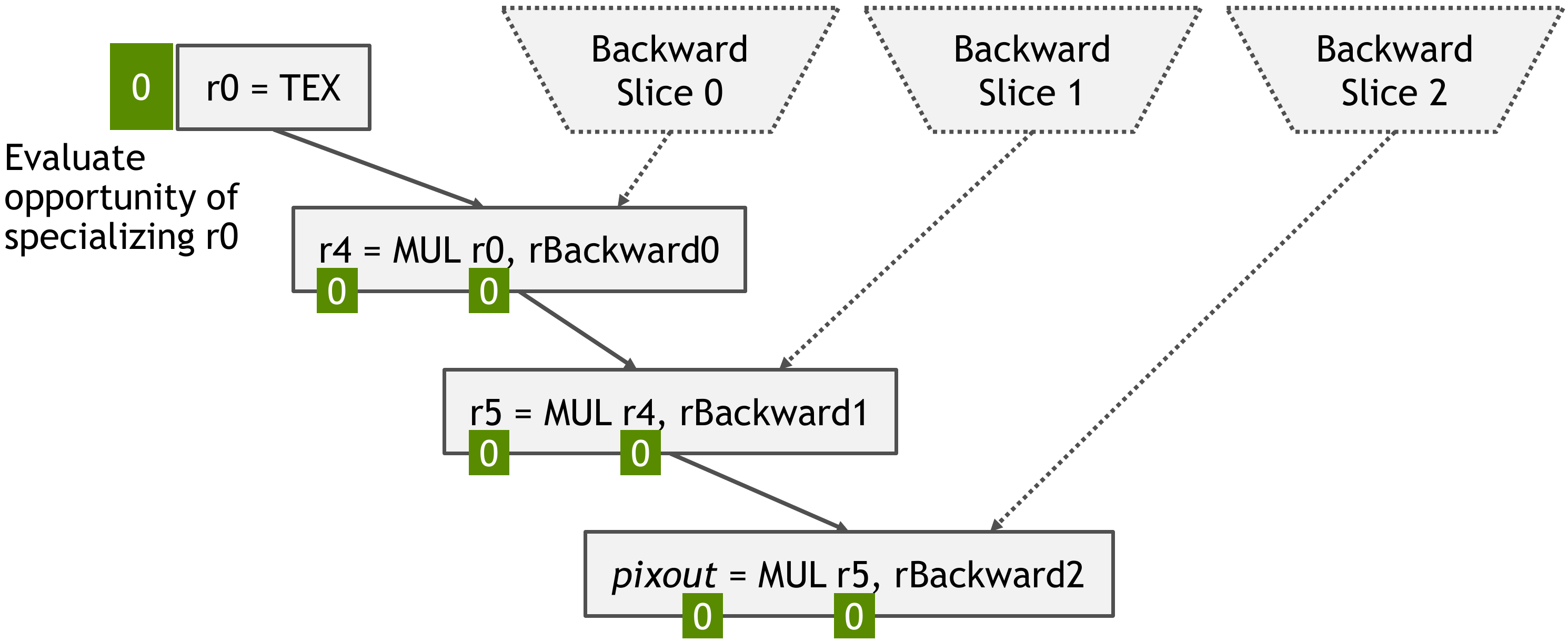}
\caption{Evaluating a candidate.  Constant propagation and folding can convert some instructions to constant literal moves.  A subsequent dead code elimination pass removes backward slices of computation that are no longer needed.}
\label{fig:evaluatinggoodness}
\end{figure}

Figure~\ref{fig:evaluatinggoodness} illustrates \autoploit's approach on a simple example. At a high-level \autoploit\ estimates benefits by first finding the set of instructions that become specialized constants in the forward slice.  In the figure, which shows the evaluation of variable \texttt{r0}, constant propagation causes all subsequent instructions to evaluate to zero. \autoploit\ then runs a dead code elimination analysis to remove the backward slices of computation that are no longer required (shown with dotted lines in the figure).  While constant propagation and dead code elimination form \autoploit's foundation, candidate estimation performs these supporting analyses:

\begin{figure*}[th!]
\centering
     \begin{subfigure}[b]{0.43\textwidth}
         \centering
         \includegraphics[width=\textwidth]{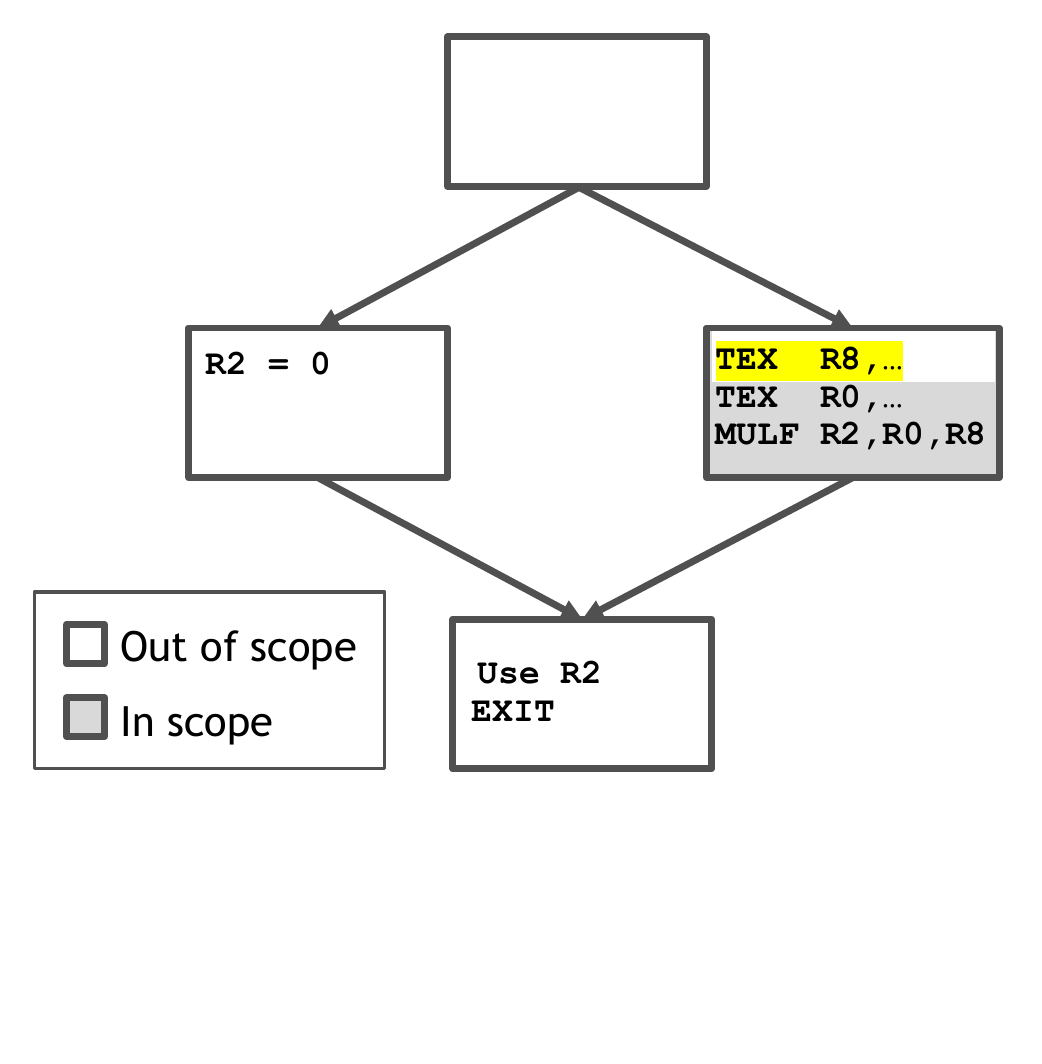}
         \caption{Original CFG}
         \label{fig:cfg_orig}
     \end{subfigure}
     \hspace{0.05\textwidth}
     \begin{subfigure}[b]{0.43\textwidth}
         \centering
         \includegraphics[width=\textwidth]{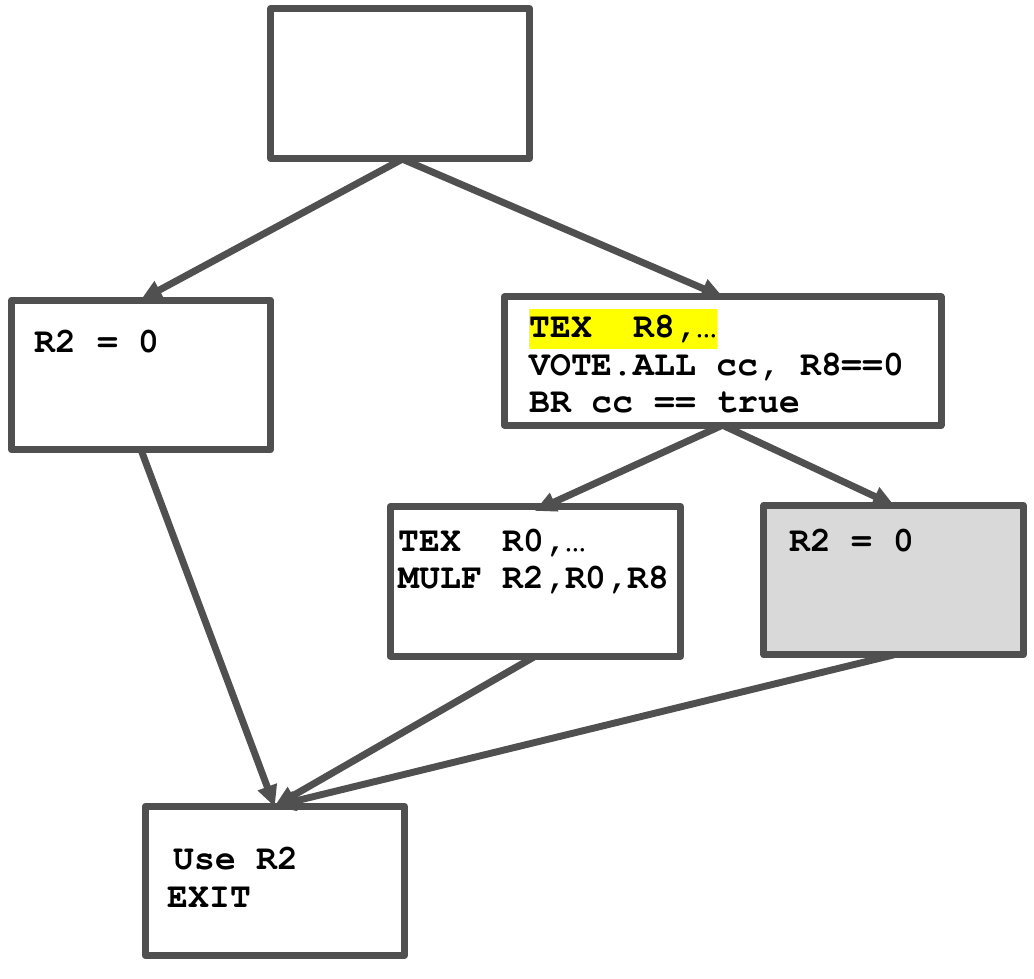}
         \caption{Transformed CFG}
         \label{fig:cfg_xformed}
     \end{subfigure}
\caption{The \autoploit\ transformation.  In this example, \autoploit\ specializes the CFG in (a) for cases when \texttt{R8} is zero (the result of the highlighted texture operation).  The specialization, shown in (b) splits the basic block in which \texttt{R8} is defined to insert a versioning check.  The {\em vote.all} instruction ensures the versioning branch is SIMT warp-convergent.}
\label{fig:thetransform}
\vspace{-0.1in}
\end{figure*}

\begin{enumerate}[wide, labelwidth=!, labelindent=0pt]
    \item {\bf Versioning check hoisting}: This step aims to determine legal, ``earlier'' locations in the program to which to hoist the computation needed to perform the candidate's versioning check.  Hoisting the versioning check can increase the scope of the computation's backward slices.  While we could have implemented a sophisticated code motion approach (\emph{e.g.}, loop invariant code motion, partial redundancy elimination, and unification), our prototype performs simple intra-block hoisting of the candidate's versioning check.  This limitation trades off missing some specialization opportunities for improved compile-time speed. 
   \item {\bf Region identification}: This step identifies the set of basic blocks, $\mathcal{R}$, in which any specialization is legal for a given candidate, and it depends on the location of the versioning check.  The set is formally defined as $\{ b~|~b \in G $ and $ v \in Dom(b) \}$, where $G$ represents the set of basic blocks in the control flow graph, $v$ is the block with the versioning check, and $Dom$ is the function that returns the \emph{dominators} of the given block.  We can copy the blocks to a new set $\mathcal{S} \gets \mathcal{R}$, and insert an if-then-else hammock that branches either to $\mathcal{R}$ or $\mathcal{S}$, depending on the outcome of the versioning check.\footnote{We perform all of the analyses for opportunity estimation \emph{without} actually transforming the shader.  This helps us improve the efficiency of our optimization.} We will specialize the instructions in $\mathcal{S}$.
   \item {\bf Constant propagation}: Within $\mathcal{S}$ we perform a standard forward constant propagation analysis.  When propagating constants, if the definition of a source operand reaches from a block $b \notin \mathcal{S}$ we conservatively assume the operand is not constant (\emph{i.e.}, it is $\top$).  This analysis discovers the set of instructions for which all destination results will provably return constants.
   \item {\bf Dead code elimination}: Source registers are not required to produce a statically known constant, and therefore newly discovered constant expressions in $\mathcal{S}$ might eliminate additional expressions in the backward slice. The following dataflow equations, which are slight modifications to live variable analysis, show how the backward analysis proceeds at the instruction level.
\begin{equation}
    Kill_i = Defs(i)
\end{equation}
\begin{equation}
    Gen_i = \begin{cases}
                \emptyset & \displaystyle\mathop{\forall}_{d \in Defs(i)} d \notin Out_i \text{ or $d$ is constant} \\
                Uses(i) & \text{otherwise} 
            \end{cases}
\label{eqn:dcegen}
\end{equation}
If all of an instruction's destination registers (\emph{i.e.}, the LHS of the instruction) are marked as constants then Equation~\ref{eqn:dcegen} does not generate any upward exposed flows.  Furthermore, if an instruction's definition is not live at the point of definition, then we can effectively remove the instruction and forgo its upward exposed flows. Of course, this only applies to instructions that do not change global state such as branches and stores. In addition, we logically treat fused-multiply-add instructions as a multiply followed by an add, which allows us to remove backward slices of the ``other'' multiplicand, when one of the multiplicands is zero. 

At the block level, Equation~\ref{eqn:dcein} is exactly the same as it is for live variable analysis.  However, we slightly modify the live variable analysis flow in Equation~\ref{eqn:dceout} so that we need only consider computing flows in the region $\mathcal{S}$.  This relies on a pre-computed set of ``live in'' virtual registers for the original shader, $LIn$.
\begin{equation}
In_n = (Out_n - Kill_n) \cup Gen_n 
\label{eqn:dcein}
\end{equation}
\begin{equation}
Out_n = \bigcup_{s \in succ(n)} \begin{cases}
                                    In_s & s \in \mathcal{S} \\
                                    LIn_s & s \notin \mathcal{S}
                                \end{cases}
\label{eqn:dceout}
\end{equation}
   \item {\bf Region refinement:} Remember that the versioning variable's basic block dominates $\mathcal{S}$.  However, constant propagation and dead code elimination may have found specialization opportunities in only a subset of $\mathcal{S}$'s basic blocks.  This optional step attempts to limit the scope of the versioning variable to reduce the number of new basic blocks specialization introduces.
\end{enumerate}

\noindent
{\bf Heuristic:} After we have run the analyses mentioned above we estimate the candidate's opportunity.  The heuristic we use to determine a candidate's effectiveness follows:
\begin{equation}
    \begin{gathered}
    T_{\autoploit} = P({vr}_c = 0) \cdot T_{\mathcal{S}} + (1 - P({vr}_c = 0)) \cdot T_{\mathcal{R}} + T_{check} \\
    T_\delta = T_{\autoploit} - T_{\mathcal{R}}
    \end{gathered}
\end{equation}
$T_{check}$ is an estimate for the number of cycles spent performing the versioning check, $T_{\mathcal{R}}$ estimates how many dynamic cycles are spent in $\mathcal{R}$, and is based on the basic block frequencies (from the profile) as well as the mixture of instructions in $\mathcal{R}$.  Likewise, $T_{\mathcal{S}}$ is a cycle estimate using basic block frequency information as well as the set of instructions remaining in $\mathcal{S}$. Intuitively, $T_\delta$ is the number of cycles \emph{saved} via specialization.  Initially we transformed any candidate if $T_\delta$ exceeded 25 cycles, but that approach failed for a couple of reasons.  First, it did not take into account the speedup of specializing the candidate.  For example, if $\mathcal{R}$ is large and expensive to execute, then saving 25 cycles is probably not worth the negative effects of \autoploit's transformation.  In addition, we want to forgo transforming candidates that introduce too many additional basic blocks.  Finally, we give an extra boost to candidates where specialization removes memory operations:
\begin{equation}
    \begin{gathered}
    su = \frac{T_{\mathcal{R}} - (T_{\mathcal{S}} + T_{check})}{T_{\mathcal{R}}} \\
    su_{thresh} = e^{\frac{-|\mathcal{S}|}{50 + (10 \cdot MemKilled}},
    \end{gathered}
\end{equation}
\noindent
where $|\mathcal{S}|$ is the number of blocks in $\mathcal{S}$, and $MemKilled$ is the number of memory operations in $\mathcal{R}$ that are not in $\mathcal{S}$.  We will not transform any candidate unless $T_\delta > 25$ \emph{and} $su < su_{thresh}$.  These refinements to our heuristic avoid transforming some empirically negative candidates.  When discussing results in Section~\ref{sec:eval}, we show that this heuristic effectively identifies several outstanding opportunities, but we also show that there are still cases where our heuristic fails.  Future work will consider alternative heuristics.

\subsubsection{Performing the transformation}

Figure~\ref{fig:thetransform} provides a simple example of the \autoploit\ transformation.  In the example, we specialized the CFG in Figure~\ref{fig:cfg_orig} for executions where \texttt{R8} is zero.  While the transformation is conceptually simple, this example lets us recap some interesting tradeoffs. As during the evaluation phase, we determine the scope of the candidate's versioning variable.  In our example, the grayed basic block is the candidate's scope, since this block is the only block in the CFG that the candidate dominates.  Our transformation creates the region $\mathcal{S}$ (the grayed block in Figure~\ref{fig:cfg_xformed}), and sets all occurrences of \texttt{R8} in $\mathcal{S}$ to zero.  It inserts a versioning check that either branches to $\mathcal{S}$ or the original region $\mathcal{R}$. \autoploit\ passes the versioning check predicate through the {\em vote.all} operation~\cite{ptx:19}, thus making the dependent branch warp-convergent. This avoids SIMT warp-divergence penalties in the transformed code. At this point, \autoploit\ invokes constant folding and dead code elimination on the shader. 

This example makes clear that the \autoploit\ transformation increases the size and complexity of a shader's CFG.  In addition, the runtime versioning check introduces overheads.  In this example, depending on $P(R8 = 0)$, the added complexity may be worthwhile because specialization removed an expensive texture lookup (\texttt{TEX}).  Note however, that specialization also \emph{delays} the execution of the ``\texttt{TEX R0 ...}" instruction.  In the original CFG in (a), the two texture operations could potentially execute concurrently, whereas specialization defers the execution of a texture operation when $R8 \neq 0$. These tradeoffs are adequately captured by various parameters used in our heuristic.   

Finally, Figure~\ref{fig:cfg_orig} demonstrates different scoping alternatives.  In \zeroploit\ the authors discuss hoisting the versioning variable as ``early" as possible, which might expose additional opportunities.  Thus the \zeroploit\ approach, which we remind the reader was not automated, would presumably hoist the versioning variable to the grayed block's dominator, where the subsequent analysis would determine that \texttt{R2} is zero everywhere when \texttt{R8} is zero.  

\subsubsection{Compile time considerations}

We have previously mentioned that our prototype does not consider inter-block code motion to increase a candidate's scope, and we made this decision primarily in the name of compile-time complexity.  Partial redundancy elimination and related techniques are expensive enough that some compiler engineers are reluctant to apply them once~\cite{khedker2017data}, let alone potentially \emph{hundreds} of times, once per candidate.

Later, in Section~\ref{sec:eval}, we show the correlation between compile time and the number of candidates \autoploit\ considers and the number of transforms it performs (up to \texttt{xform\_limit} in Figure~\ref{fig:zp_compiler_flow}).  We can reduce compilation time by better candidate filtering, including increasing the ``probability of zero" threshold for a candidate.  Though we have not done so, an implementation could ignore testing scalar candidates in which the candidate's versioning variable is also part of a vector candidate.

\section{Evaluation}
\label{sec:eval}

This section describes our experimental methodology, pre\-sents runtime speedps, compile-time overheads, and concludes with a discussion of some interesting insights regarding the results.

\subsection{Methodology}

Table~\ref{tab:applist} lists the gaming applications evaluated in this paper. Using an internal frame-capture tool, similar to publicly available tools like Renderdoc~\cite{renderdoc}, Nsight~\cite{nsight}, etc., one or more random single-frames are captured from a built-in benchmark run or from actual gameplay of each gaming application, depending on whether the game has a built-in benchmark. These single-frame  captures, called APICs, contain all the information needed to replay a game frame i.e. both the API sequence (including all relevant state) as well as shader programs used in individual calls. 
For applications for which we captured more than one APIC, we made sure to capture them from visually different scenes. We evaluate \autoploit\ on a total of 14 APICs spanning 8 applications as shown in Table~\ref{tab:applist}. We hyphenate an application's short name with the APIC number to uniquely refer to an APIC in this section (for e.g., \app{SS-3} to refer to the third APIC of Serious Sam 4).

\begin{table}[h!]
  \begin{center}
    \caption{Gaming applications evaluated in this paper.}
    \label{tab:applist}
\begin{scriptsize}
     \begin{tabular}{|c|c|c|c|c|c|}
      \hline
      \textbf{Application} & \textbf{Short} & \textbf{APICs} & \textbf{dx11} & \textbf{dx12}  \\
    \hline
     Ashes of the Singularity - Escalation & Ashes & 3 & \checkmark & \\
     \hline
    Deus Ex Mankind Divided & DXMD &    1 &  & \checkmark  \\
     \hline
     Final Fantasy XV  & FFXV & 1 &  \checkmark & \\
     \hline
     Metro Exodus & Metro &  1 &  \checkmark &   \\
    \hline
      PlayerUnknown's Battlegrounds & PUBG  & 3 & \checkmark &  \\
    \hline
      Horizon Zero Dawn & HZD  &  1 &  & \checkmark  \\
      \hline
      WatchDogs Legion & Watch & 1 &  & \checkmark  \\
      \hline
      Serious Sam 4 & SS & 3 & \checkmark & \\
      \hline
    \end{tabular}
\end{scriptsize}
  \end{center}
\end{table}

We implemented \autoploit\ as a research prototype on top of a recent branch of \nvidia's GeForce Game Ready driver.  
Value profiling is performed in dedicated offline runs, while just-in-time compilation based on these value profiles happens online. 

We perform our experiments on an \nvidia\ \gpuname\ GPU, locked to base clock settings of 1515 MHz for the GPU core and production DRAM frequency settings. Full specification of this GPU can be found here~\cite{rtx2080}.  We compare the final rendered image against a golden reference image to ensure that our transforms are functionally correct. We measure GPU-only time using accurate in-house profiling tools, after locking GPU clocks and power state to production active game-play settings. These runs use production settings for driver and compiler optimizations, which include classical optimizations like constant folding, dead code elimination, loop unrolling, instruction scheduling, etc., in addition to various machine-specific optimizations. 

\subsection{Shader Program Performance}

\begin{figure}
\centering
\begin{subfigure}{\linewidth}
\includegraphics[width=\linewidth]{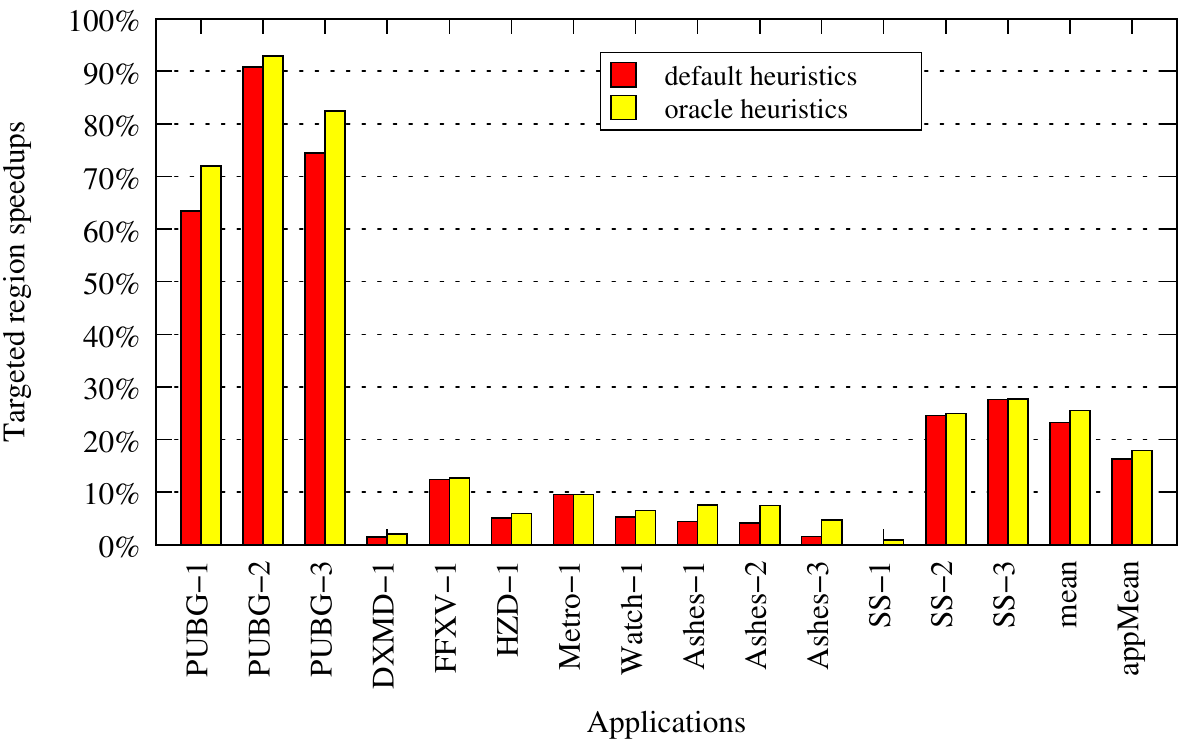}
\caption{Speedup for targeted regions with default and oracle heuristics.}
\end{subfigure}
\begin{subfigure}{\linewidth}
\includegraphics[width=\linewidth]{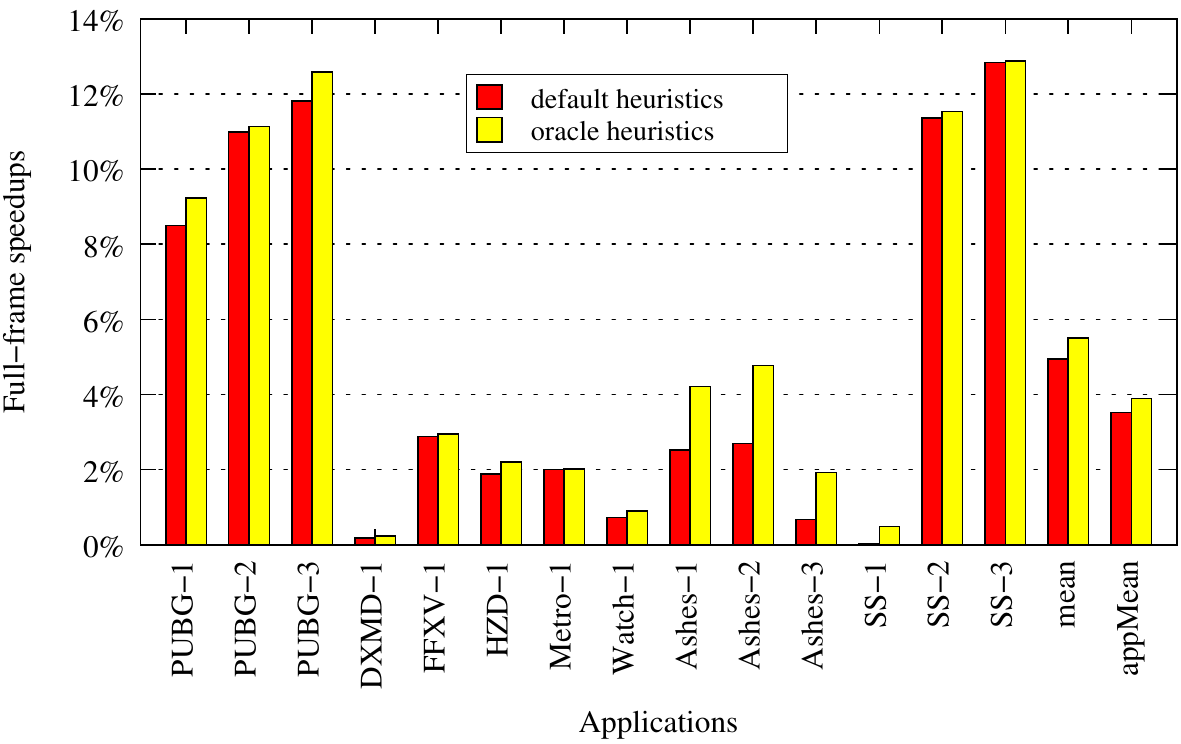}
\caption{Full-frame speedups with default and oracle heuristics.}
\vspace{-0.1in}
\end{subfigure}
\caption{Performance of \autoploit\ on modern gaming applications.}
\label{fig:speedups}
\vspace{-0.2in}
\end{figure}

\autoploit's speedups over baseline execution at the targeted region and full-frame levels are shown in Figure~\ref{fig:speedups}a and Figure~\ref{fig:speedups}b, respectively. We define a ``targeted region'' of an application as the set of shader programs that \autoploit's heuristic chooses for transformation. We calculate targeted region speedup as the sum of the baseline time for all chosen shaders divided by the sum of the execution times for the same shaders in an \autoploit-transformed run. Both graphs show performance speedup with default \autoploit\ heuristics as well as an ``oracle'' heuristic. The latter is calculated by simply ignoring all heuristic-induced slowdowns. 

Since our goal in this paper is to demonstrate an automatic compiler technique for \zeroploit\ and not to explore its effectiveness  across scenes of gaming applications, we use the same APICs for profiling and testing. However, in the few applications for which we were able to obtain multiple APICs corresponding to different scenes, we can see that \autoploit\ is able to extract \zeroploit\ benefit effectively across scenes. For example, in Figure~\ref{fig:speedups}a, notice similar upsides among \app{PUBG} APICs, among \app{Ashes} APICs, and in APICs 2 and 3 of \app{SS}. This augurs well for a dynamic adaptive compilation system based on \autoploit, which we plan to explore in the future.

At the full-frame level, we see average speedups of \fpsSpeedup\ and \fpsBestOfSpeedup\, with default and oracle heuristics, while for targeted regions \autoploit\ achieves average speedups of \avgBucketSpeedup\ and \avgBestOfBucketSpeedup, with default and oracle heuristics, respectively. From these graphs, we can see that the default \autoploit\ heuristics are able to effectively target \zeroploit\ opportunities in most applications. 

\subsection{Compilation Statistics}

Although \autoploit\ is reserved for compiling hot shaders, it does run in the context of a just-in-time compiler where compilation time is a consideration.  To determine the overhead of our compiler analyses and transformations we collected the compile-time slowdown over the driver's baseline compilation for each shader that \autoploit\ managed to transform.  We focus on the transformed shaders to guarantee that the evaluate-transform loop executes at least once.  We use a proprietary tool to measure and record the Direct3D driver's per-shader compile times.

On a 3.4Ghz Intel Core i7-6800K CPU with 32 GB of RAM, the geometric mean compile-time slowdown was $1.57\times$ over 138 varied shaders across all of our applications.  If we increase the ``probability of zero" threshold ($P_{thresh}$) for candidates that \autoploit\ considers from $0.32$ to $0.9$, the mean slowdown decreases slightly to $1.45\times$.  If, on the other hand, we limit the number of iterations of the evaluate-transform loop to one, we see the mean slowdown decrease to $1.35\times$.  Not surprisingly, these results show a clear correlation between the number of candidates evaluated and the resultant overhead.  Of note, modern drivers carefully chose when to asynchronously compile optimized shaders and will not waste CPU cycles compiling additional shaders unless performance is limited by the GPU.  In such cases, where the CPU is not limiting performance, spending additional time optimizing shaders can be prudent.

\subsection{Discussion}

\subsubsection{Heuristic Performance}

\begin{figure}
\centering
\includegraphics[width=\linewidth]{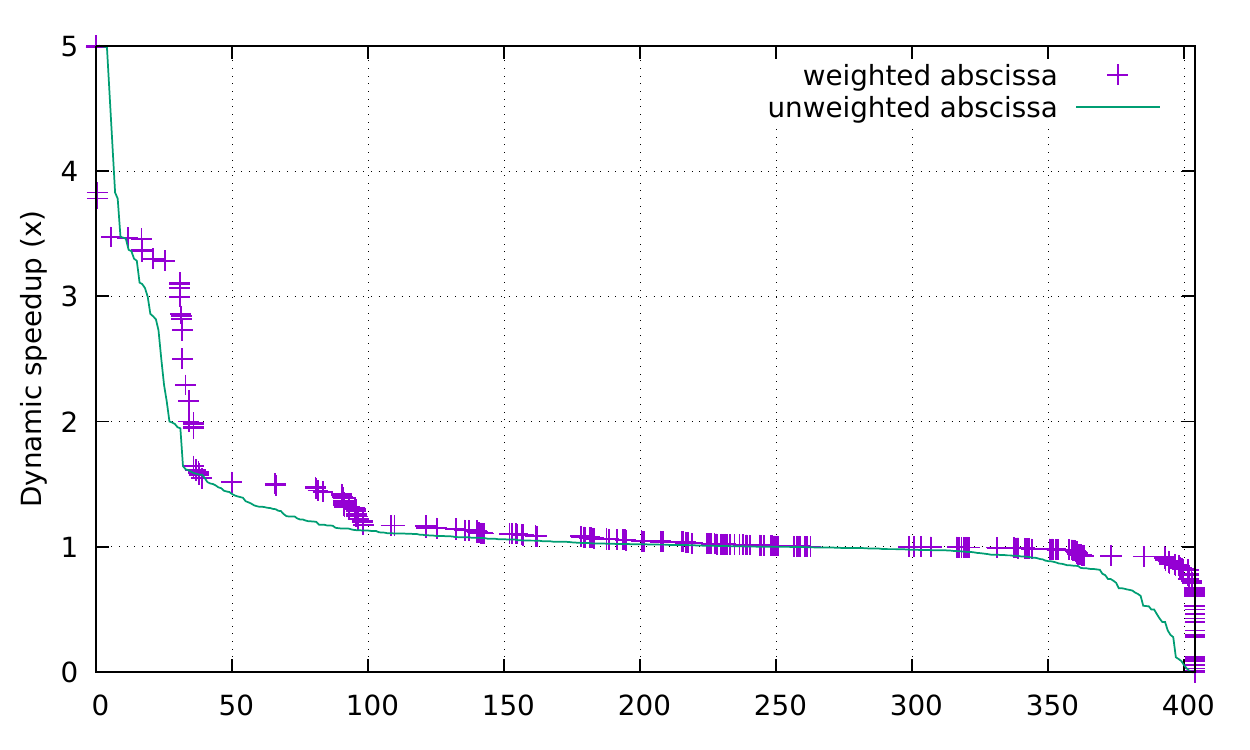}
\caption{Performance of 404 shaders across our test suite that \autoploit's heuristic deemed worthy of transforming.}
\label{fig:heuristicperf}
\vspace{-0.1in}
\end{figure}

Figure~\ref{fig:heuristicperf} shows how \autoploit's heuristics performed at the individual shader level. Here, we plot individual shader speedups on the y-axis against both an unweighted abscissa (blue line) and a weighted abscissa (magenta points). The unweighted abscissa is simply a monotonically increasing shader identifier, whereas the weighted abscissa is calculated based on a given shader's normalized frame-time contribution in its baseline execution. We can see that \autoploit\ is positive or neutral in most shaders. On about 25\% of the shaders, it produces slowdowns as can be seen on the right end of blue line. However, as the steep fall of the cluster of magenta points at the right end shows, these slowdowns predominantly occur in shaders with negligible frame-time contributions. This implies \autoploit's heuristics work well in practice on shader programs that matter. 

Through manual inspection of a few of the negative examples in Figure~\ref{fig:heuristicperf} we found that \autoploit's code-versioning transform at times interacts pathologically with downstream passes. \autoploit\ currently runs early in the compiler flow, before loop unrolling, scheduling, and register allocation.  We have seen examples where \autoploit\ inhibits unrolling an important loop, and we have seen it negatively affect texture scheduling and register usage. As future work, we plan to investigate ways to avoid such cases through heuristic enhancements.

\subsubsection{Effectiveness of \autoploit\ in {\tt PUBG}}
PUBG contains an interesting, positive example.  \autoploit\ progressively optimized an important shader over three iterations of the evaluate-transform loop.  In the first iteration, the best candidate removed $30$ multiplies and several multiply accumulate and add instructions.  The candidates that \autoploit\ transformed on the second and third iteration scored poorly on the first iteration; the first transformation exposed opportunities that did not exist.  On the second iteration a vector candidate removed dozens of additional math operations.  Again, the final and most influential candidate that was transformed scored poorly on the second iteration.  On the final iteration of the evaluate-transform loop \autoploit\ removed a large swath of instructions, including dozens of memory operations.

\section{Related work}
\label{sec:related}

\zeroploit, a profile-guided code transform for forward and backward slice specialization based on zero values, was shown to be effective in gaming applications with manually optimized codes~\cite{rangan:20:taco}. In this paper, we highlight the various challenges that need to be solved in order to automate \zeroploit, and present and evaluate a full compiler solution for it. 

Value-dependent code specialization has been previously automated in partial evaluation systems for functional and imperative language programs~\cite{jones:93:book, consel:96:ispe},  general purpose programs~\cite{calder:99:jilp, muth:00:sas, grant:99:pldi, sastry:00:fddo}, embedded software~\cite{chung:02:cadics}, Java just-in-time compilers~\cite{shankar:05:oopsla, costa:13:cgo}, etc. 
Unlike aforementioned value-specialization efforts, which specialize for generic values and thus perform only forward-slice specialization, \autoploit\ achieves both forward and backward-slice specialization by specializing for just zeros. For example, Muth et al.'s code-versioning approach for generic value specialization used heuristics that estimate savings from specializing forward slices~\cite{muth:00:sas}. Grant et al. used an annotations-based system without heuristics to target dynamic zero candidates at function scopes in the {DyC} dynamic compilation system~\cite{grant:99:pldi}. In contrast, \autoploit\ estimates forward and backward slice specialization benefits by using expected savings from constant propagation and folding as well as dead-code elimination.

Recently, Leobas and Pereira identified the type of opportunity targeted by \zeroploit\ as the mathematical notion of {\em semirings}. As a proof-of-concept semiring optimization, they automated a profile-driven transform to target silent stores and evaluated it on CPU programs in the LLVM test suite~\cite{leobas:20:oopsla}. They also presented a novel online profiler that works well for detecting silent stores in hot loops. \autoploit\ differs from semiring optimization in a few key points. First, \autoploit's focus on gaming applications means it can take advantage of developer-granted permission to perform IEEE-754-unsafe operations to optimize floating point code. Second, \autoploit\ uses a cost-benefit analysis based on constant folding and dead code elimination to evaluate, rank, and select from among several versioning variables per shader program, whereas the semiring silent store optimization did not have to use any heuristics and was applied to all stores detected by their profiler as being likely silent. Third, their loop-iteration sampling based profiler cannot be directly applied to graphics programs, since most shader programs do not have loops or ones that iterate more than a few times. Therefore, \autoploit\ uses a profiling strategy more suited to gaming applications - instruction granular zero-value profiling, sampled over several frames. And fourth, we describe the \autoploit\ compiler pipeline in full detail, complete with dataflow equations, which complements Leobas and Pereira's theoretical proofs for semiring optimizations, thus enabling provably correct practical application of \zeroploit\ and similar semiring optimizations.

Finally, unlike prior work, \autoploit, when applied to gaming applications, has to contend with the SIMT execution model of \nvidia\ and similar GPUs. \autoploit\ accomplishes this by using \nvidia's warp-wide {\em vote} instruction~\cite{ptx:19} in both its zero-value-profiler as well as its code transformer.

\section{Conclusion}
\label{sec:conc}

Recent work called \zeroploit\ identified a profile-guided optimization that exploits certain dynamically zero valued operands in graphics shader programs to execute highly specialized code. However, this prior work relied on manual effort to identify opportunities and transform them. In this paper, we advanced the above line of work with \autoploit, which uses novel compile-time heuristics to automatically identify opportunities and transform code. With its default heuristics, \autoploit\ achieves an average frame-rate improvement of \fpsSpeedup\ and an average targeted region speedup of \avgBucketSpeedup\ over a heavily optimized production driver, across a suite of modern gaming applications.

Our current results based on single-frame testing are encouraging.  Future work will study \autoploit's effectiveness both as a PGO based on offline profiling as well as an adaptive compilation strategy based on continuous online profiling, in multi-frame scenarios such as demos and game-play.

\bibliographystyle{ACM-Reference-Format}
\bibliography{paper}


\end{document}